\documentstyle[12pt]{article}
\begin{document}

\renewcommand{\baselinestretch}{1.5}
\newcommand\beq{\begin{equation}}
\newcommand\eeq{\end{equation}}
\newcommand\bea{\begin{eqnarray}}
\newcommand\eea{\end{eqnarray}}

\centerline{\bf Applications of the Collective Field Theory for the}
\centerline{\bf Calogero-Sutherland Model}
\vskip 1 true cm

\centerline{Diptiman Sen \footnote{Permanent Address: Centre for Theoretical
Studies, Indian Institute of Science, Bangalore 560012, India} and   
R. K. Bhaduri}
\small
\centerline{\it Department of Physics and Astronomy, McMaster University,}
\centerline{\it Hamilton, Ontario L8S 4M1, Canada}
\vskip .25 true cm
\normalsize
\vskip 2 true cm

\noindent 
{\bf Abstract}
\vskip .5 true cm
We use the collective field theory known for the Calogero-Sutherland model
to study a variety of low-energy properties. These include the ground state 
energy in a confining potential upto the two leading orders in the particle 
number, the dispersion relation of sound modes with a comparison to the two 
leading terms in the low temperature specific heat, large amplitude waves, 
and single soliton solutions. The two-point correlation function derived from 
the dispersion relation of the sound mode only gives its nonoscillatory 
asymptotic behavior correctly, demonstrating that the theory is applicable 
only for the low-energy and long wavelength excitations of the system.

\vskip 1 true cm
\noindent PACS numbers: ~03.40.Kf, ~03.65.Ge
 
\newpage

\centerline{1. INTRODUCTION}
\vskip .5 true cm

The Calogero-Sutherland-Moser model (CSM) [1-3] has attracted 
much attention in recent years due to its relation to a wide variety of 
interesting problems. Some examples are random matrix theory \cite{MEH},
quantum spin chains with long-range interactions \cite{HAL1}, generalized
exclusion statistics [6-12], Gaussian 
conformal field theories \cite{FUK}, edge states in a quantum Hall system
\cite{YU}, and nonlinear internal waves in a stratified fluid \cite{CHE}. 

The CSM has been known to be exactly solvable 
and integrable, both classically and quantum mechanically, for quite some 
time [1-3,~16]. However detailed investigations into its 
collective properties have begun only recently [17-19]. 
A collective field theory to study the excitations of a superfluid, as well as 
the ground state of a condensed Bose-Einstein gas was developed long 
back \cite{BEC}. In such a theory, the fundamental coordinate is the 
density field [21-24]. For the CSM,  
the results obtained so far include the ground state energy of the model 
placed in a harmonic oscillator potential, waves of arbitrary amplitude for 
strong coupling, and isolated solitons on an uniform background density. 

In our paper, we will study essentially the same aspects but in 
more detail and for arbitrary coupling, thereby generalizing the earlier
results in several ways. Wherever appropriate, we will compare 
our results with those obtained earlier by other methods \cite{ISA2,MUC}.
This will illustrate that certain properties of the model can be derived more 
easily and generally from collective field theory. These properties include
the dispersion relation for small amplitude and long wavelength sound modes,
the low temperature specific heat, and the two-point correlation function. 
The collective field theory yields a dispersion relation for the sound mode 
that terminates exactly in the second order of the wave number, which is 
adequate only for small wave numbers. This gives the correct {\it 
nonoscillatory} behavior of the two-point correlation function for 
asymptotically large distances, and it fails at shorter distances. The 
collective field theory formulation is thus seen to be a useful description 
for the low-energy, or small wave number excitations of the CSM system.  

\vskip .5 true cm
\centerline{2. CALOGERO-SUTHERLAND MODEL AND COLLECTIVE}
\centerline{FIELD THEORY}
\vskip .5 true cm

The simplest form of the CSM consists of particles on a line which interact 
pairwise through an inverse-square potential. The model can also be defined 
on a circle with periodic boundary condition \cite{SUT1}; the two versions 
of the model have identical physical properties in the thermodynamic limit 
in which the the number of particles $N$ and the length $L$ of the line (or 
circle) are simultaneously taken to infinity keeping the particle density 
$\rho_0 =N/L$ fixed. The Hamiltonian for particles on a line is given by
\beq
H ~=~ \sum_{i=1}^N ~\frac{p_i^2}{2m} ~+~ \frac{\hbar^2 \lambda (\lambda -
1)}{m}~ \sum_{i<j} ~\frac{1}{(x_i - x_j)^2} ~,
\label{hamcsm}
\eeq
where the dimensionless coupling $\lambda \ge 0$. To make the problem
well-defined quantum mechanically, we have to add the condition that
the wave functions $\Psi$ goes to zero as $\vert x_i - x_j \vert^{\lambda}$
whenever two particles $i$ and $j$ approach each other. For $\lambda = 0$
and $1$, the model describes free bosons and free fermions respectively. 
Since the two-body potential is singular enough to prevent particles from
crossing each other, we can choose the wave functions to be either symmetric 
(bosonic) or antisymmetric (fermionic). The energy spectrum is the same in 
the two descriptions.

Let us briefly summarize some of the exactly known results for this model
\cite{SUT1,SUT2}. If $E_0$ denotes the ground state energy, then the
chemical potential at zero temperature is given by $\mu = \partial E_0 /
\partial N$ in the thermodynamic limit. This takes the form
\beq
\mu ~=~ \frac {\pi^2 \hbar^2 \lambda^2 \rho_0^2}{2m} ~.
\label{mu}
\eeq
In a fermionic description of the model, it is natural to define a Fermi 
momentum 
\beq
p_0 ~=~ \pi \hbar \rho_0 ~.
\eeq
(We should point out that some papers in this field find it more convenient 
to define the Fermi momentum to be $\pi \hbar \lambda \rho_0$).
The low-energy excitations of (\ref{hamcsm}) are known in detail. They can 
be thought of as being made up of particle and hole excitations. Let us first
define the sound velocity by the relation 
\beq
v_s^2 ~=~ \frac{\rho_0}{m} ~\left( \frac{\partial \mu}{\partial \rho_0} 
\right) ~.
\eeq
Then
\beq
v_s ~=~ \frac{\pi \hbar \lambda \rho_0}{m} ~.
\label{vs}
\eeq
It is known that the particle excitations necessarily have $\vert p \vert 
\ge p_0$, with the dispersion
\beq
\epsilon_p (p) ~=~ \frac{1}{2m} ~(\vert p \vert - p_0 ) ~[~ \vert p \vert + 
(2 \lambda -1) p_0 ~] ~+~ \mu ~.
\label{enp}
\eeq
The hole excitations have $\vert p \vert \le p_0$, with the dispersion
\beq
\epsilon_h (p) ~=~ \frac{\lambda}{2m} ~(p_0^2 - p^2) ~-~ \mu ~.
\label{enh}
\eeq
If we define group velocities $v = \hbar \partial \omega / \partial p$, we 
find that particles have $\vert v \vert \ge v_s$, while holes have $\vert 
v \vert \le v_s$. The sound velocity in (\ref{vs}) can be obtained by 
considering a sound mode to be made up of a particle with energy-momentum 
$(\epsilon_1, p_1)$ and a hole with energy-momentum $(\epsilon_2, p_2)$. 
Then $v_s = (\epsilon_1 + \epsilon_2)/(p_1 - p_2)$ in the limit $p_1, p_2
\rightarrow p_0$.

The collective field theory for the CSM is obtained by changing variables from 
the particle coordinates $x_i$ to the density field $\rho (x)$ defined as
\beq
\rho (x) ~=~ \sum_i ~\delta (x-x_i) ~.
\label{rho}
\eeq
As emphasized in reference \cite{JEV1}, such a change of variables is 
meaningful
only if the particle number $N \rightarrow \infty$. We therefore have to check 
at various stages whether the results obtained from collective field theory 
for {\it finite} values of $N$ are indeed correct for the model defined in 
(\ref{hamcsm}). For this reason, we will compare the collective field 
theory results with those obtained by other methods whenever possible.

After changing variables, the quantum Hamiltonian takes the form 
\cite{AND2,AND3}
\bea
H ~=~ \frac{\hbar^2}{2m} ~\int dx ~&\Bigl[ & \rho ~(\partial \theta)^2 ~
+~ \frac{\pi^2 \lambda^2}{3} ~\rho^3 \nonumber \\
&& \quad +~ \lambda (\lambda -1) ~\rho_H \partial \rho ~+~ \frac{(\lambda 
-1)^2}{4} ~\frac{(\partial \rho )^2}{\rho } ~\Bigr] ~, 
\label{hamcft}
\eea
where $\partial = \partial / \partial x$, and $\rho_H$ is proportional to the 
Hilbert transform of $\rho$ defined as the principal part integral \cite{HIL},
\bea
\rho_H (x) ~&=&~  ~\int dy ~\rho (y) ~\frac{P}{x-y} ~, \nonumber \\
{\rm where} \quad \frac{P}{x-y} ~&\equiv & \frac{1}{2} ~\lim_{\epsilon 
\rightarrow 0} ~(~ \frac{1}{x-y+i \epsilon} ~+~ \frac{1}{x-y -i \epsilon} ~)~.
\eea
The field $\hbar \theta$ is canonically conjugate to $\rho$, and they satisfy 
the equal-time commutation relation 
\beq
[ ~\rho (x) ~,~\hbar \theta (y) ~] ~=~ i \hbar ~\delta (x-y) ~.
\label{comm}
\eeq
We may therefore set $\theta (x) = - i \delta / \delta \rho (x)$, and try to
find eigenstates of the Hamiltonian (\ref{hamcft}) as functionals of $\rho 
(x)$. Although this is sometimes possible \cite{AND3}, it is generally very 
difficult to find exact eigenstates.

We will therefore take the simpler route of studying 
(\ref{hamcft}) {\it classically}. For this purpose, it is useful to rewrite
the collective theory in terms of a fluid. Let us introduce a complex field
\beq
\psi (x,t) ~=~ \sqrt{\rho (x,t)} ~e^{i \theta (x,t)} ~.
\eeq
We define a Lagrangian $L = \int dx {\cal L}$, where the Lagrangian density
\bea
{\cal L} ~&=&~  \frac{i \hbar}{2} ~(\psi^{\star} \dot{\psi} ~-~ 
\dot{\psi^{\star}} \psi ) ~-~ \frac{\hbar^2}{2m} ~\partial \psi^{\star} 
\partial \psi ~-~ U [\rho (x) ] \nonumber \\
&=&~  -~ \hbar \rho \dot{\theta} ~-~ \frac{\hbar^2}{2m} ~\Bigl[ ~
\frac{(\partial \rho )^2}{4 \rho} ~+~ \rho (\partial \theta)^2 ~\Bigr] ~-~ 
U [ \rho ] ~, \nonumber \\
{\rm where} \quad U [ \rho ] ~&=&~  \frac{\hbar^2}{2m} ~\Bigl[ ~\frac{\pi^2 
\lambda^2}{3} ~\rho^3 ~ +~ \lambda (\lambda -1) ~ \rho_H \partial \rho ~
+~ \frac{\lambda (\lambda -2)}{4}~ \frac{(\partial \rho)^2}{\rho} \Bigr] ~,
\nonumber \\
&& 
\label{lag}
\eea
and a dot denotes $\partial / \partial t$. From (\ref{lag}) we see that 
$\rho$ and $\hbar \theta$ are canonically conjugate to each other, and we 
can recover the Hamiltonian (\ref{hamcft}) by the usual methods. 

It is interesting to note that the Lagrangian (\ref{lag}) is quadratic 
in $\psi$ and $\psi^{\star}$, and is therefore noninteracting for 
$\lambda =0$ (free bosons). This is understandable because the collective 
field theory is a bosonic theory, as is clear from (\ref{comm}). 
On the other hand, the collective theory is interacting for $\lambda =1$ 
(which is free in terms of a {\it fermionic} theory).

We now proceed to study the theory (\ref{lag}) classically. At a classical 
level, Eq. (\ref{hamcft}) (for the static case) may be regarded as the 
energy density functional of the single-particle density $\rho(x)$. This is 
analogous to the density functional theory of a correlated many-particle 
system, a highly sucessful formalism in many branches of physics \cite{SKY}.  
Since we wish 
to study the system in an external potential $V(x)$ in Section 3, let us add 
\beq
\int dx ~[~\mu - V(x)~] ~\rho (x) ~-~ \mu N
\eeq
to the Lagrangian, where $\mu$ is the chemical potential. The Euler-Lagrange 
equations of motion obtained by extremizing the action $S = \int dt L$ are 
given by
\bea
\frac{\pi^2 \hbar^2 \lambda^2}{2m} ~\rho^2 ~&-&~ \frac{\hbar^2 \lambda (
\lambda - 1)}{m} ~ \partial \rho_H ~+~  \frac{\hbar^2 (\lambda -1)^2}{8m} ~
\Bigl[ ~ \Bigl( \frac{\partial \rho}{\rho} \Bigr)^2 ~-~ 2 \frac{\partial^2 
\rho}{\rho} ~ \Bigr] \nonumber \\
&+&~ \hbar \dot{\theta} ~+~ \frac{\hbar^2}{2m} ~(\partial \theta)^2 ~-~ \mu ~
+~ V(x) ~=~ 0 ~, 
\label{eom1}
\eea
and
\beq
\dot{\rho} ~+~ \frac{\hbar}{m} ~\partial ( \rho \partial \theta) ~=~ 0 ~.
\label{eom2}
\eeq
In addition, the density must satisfy the constraint
\beq
\int dx ~\rho ~=~ N ~.
\label{cons}
\eeq
Eq. (\ref{eom2}) will be recognized as the equation of continuity since 
$(\hbar /m) \partial \theta$ is the velocity field; this can be
seen from the expression for momentum given below.

Our system has three conserved quantities, namely, the particle number $N$, 
the momentum (if there is no external potential)
\beq
P ~=~ - ~\frac{i\hbar}{2} ~\int dx ~( \psi^{\star} \partial \psi - \partial
\psi^{\star} \psi ) ~=~ \hbar ~\int dx ~\rho ~\partial \theta ~,
\label{mom}
\eeq
and the energy
\bea
E = && \frac{\hbar^2}{2m} ~\int dx ~\Bigl[ ~\rho (\partial \theta)^2 + 
\frac{\pi^2 \lambda^2}{3} \rho^3 + \lambda (\lambda -1) \rho_H \partial 
\rho + \frac{(\lambda -1)^2}{4} ~\frac{(\partial \rho)^2}{\rho} ~\Bigr] 
\nonumber \\
&& +~ \int dx ~V (x) ~\rho ~.
\label{ene}
\eea
There are probably an infinite number of conserved quantities in addition
to the three above since our original system is integrable; however 
explicit field theoretic expressions for these other quantities are not 
known.

In the absence of an external potential $V(x)$, Eqs. (\ref{eom1}) and
(\ref{eom2}) are invariant under scaling and Galilean transformations. Under 
scaling by a factor $\alpha$, we have
\bea
\rho (x,t) ~&\rightarrow&~ \alpha ~\rho (\alpha x, \alpha^2 t) ~, \nonumber \\
\theta (x,t) ~&\rightarrow&~ \theta (\alpha x, \alpha^2 t) ~, \nonumber \\
\mu ~&\rightarrow&~ \alpha^2 \mu ~.
\eea
Under a Galilean transformation by velocity $v$,
\bea
\rho (x,t) ~&\rightarrow &~ \rho (x -vt, t) ~, \nonumber \\
\theta (x,t) ~&\rightarrow &~ \theta (x -vt, t) ~+~ \frac{mv}{\hbar} ~(x- 
\frac{1}{2} vt) ~, \nonumber \\
\mu ~&\rightarrow &~ \mu ~.
\label{gal}
\eea
Thus
\bea
P ~&\rightarrow &~ P ~+~ m N v ~, \nonumber \\
E ~&\rightarrow &~ E ~+~ P v ~+~ \frac{1}{2} m N v^2 ~.
\eea

It is quite remarkable that if $V=0$, all values of $\lambda > 1$ are 
equivalent to each other according to Eqs. (\ref{eom1}) and (\ref{eom2}). 
Namely, if we redefine
\bea
\tilde{x} ~&=&~ x ~, \quad \tilde{t} ~=~ t (\lambda -1) ~, \nonumber \\
\tilde{\rho} ~&=&~ ~\frac{\lambda \rho}{\lambda -1} ~, \quad \tilde{\theta} ~
=~ \frac{\theta}{\lambda -1} ~, \nonumber \\
\tilde{\mu} ~&=&~ \frac{\mu}{(\lambda -1)^2} ~,
\label{red}
\eea
then $\lambda -1$ can be completely scaled out of (\ref{eom1}) and 
(\ref{eom2}). 
Similarly, all values of $\lambda < 1$ (but not equal to $0$) are equivalent 
to each other; we can carry out the same redefinitions as in (\ref{red}),
followed by $\tilde{\rho} \rightarrow - \tilde{\rho}$ in order to keep 
$\tilde{\rho}$ positive. If we redefine the energy functional as $\tilde{E} =
\lambda E / \vert \lambda -1 \vert^3 $, we see from (\ref{ene}) that
\beq
\tilde{E} ~=~ \frac{\hbar^2}{2m} ~\int dx ~\Bigl[ ~\tilde{\rho} (\partial 
\tilde{\theta})^2 ~+~ \frac{\pi^2}{3} \tilde{\rho}^3 ~\pm ~\tilde{\rho}_H 
\partial \tilde{\rho} ~+~ \frac{(\partial \tilde{\rho})^2}{4 \tilde{\rho}} ~
\Bigr] ~,
\label{madelung}
\eeq
where the $\pm$ signs are for $\lambda > 1$ and $\lambda < 1$ respectively.
Thus it is sufficient to study the collective field theory for just
two values of $\lambda$, one less than $1$ and the other greater than $1$.
This property of the collective field theory clearly shows that it is a 
rather coarse description of the CSM. This is to be contrasted with the exact 
solution of the model (\ref{hamcsm}) some of whose features (for instance, 
the dynamical correlation functions \cite{HA}) are sensitively dependent on 
number theoretic properties of $\lambda$.

In passing, it may be noted that formally the scaled energy density given by 
Eq. (\ref{madelung}) is of the same form as the so-called Madelung fluid 
\cite{MAD}, which is a hydrodynamical description of the one-particle 
Schr\"{o}dinger equation. In this picture, the first term is the classical 
kinetic 
energy of the fluid, the next two represent the potential energy, and the 
last term arises from the quantum kinetic energy. The latter gives rise to 
the Bohm potential \cite{BOH} in the equations of motion. This interpretation 
also holds if there are $N$ particles in the same quantum state. The 
normalization of $\tilde{\rho}$ in Eq. (\ref{madelung}), however, is 
{\it not} $N$. To pursue this line of thought more carefully, it is necessary 
to modify transformations~(\ref{red}), and scale the 
$x$-coordinate to demand $\int d \tilde{x} \tilde{\rho} =N$. The Bohm term 
in (\ref{madelung}) remains unaffected, but the interaction terms become 
$\lambda$-dependent. We will not elaborate further along these lines.    

We will now study various solutions of the equations of motion (\ref{eom1})
and (\ref{eom2}), with an external potential in Section 3 and without an 
external potential in Sections 4-6. We are interested in two kinds of 
solutions, (a) static solutions in which $\rho$ depends only on $x$ and 
$\theta =0$ (in particular, the ground state is always of this form), and 
(b) time-dependent solutions in which $\rho$ and $\theta$ depend on both $x$ 
and $t$.

\vskip .5 true cm
\centerline{3. GROUND STATE IN AN EXTERNAL POTENTIAL}
\vskip .5 true cm

For any external potential, Eq. (\ref{eom1}) gives the exact quantum ground 
state energy and density if $\lambda =0$. In that case, let $\Psi_0 (x)$ and 
$e_0$ denote the exact one-particle ground state wave function (normalized to 
unity) and energy obtained by solving the Schr\"{o}dinger equation with a 
potential $V(x)$. Then the solution of (\ref{eom1}) and (\ref{eom2}) is given 
by
\bea
\rho (x) ~&=&~ N ~\vert \Psi_0 (x) \vert^2 ~, \nonumber \\
E_0 ~&=&~ N e_0 ~.
\eea
The question therefore is how well collective field theory does for 
{\it nonzero} values of $\lambda$.

To begin with, let us consider the case of a simple harmonic potential, with 
\beq
V (x) ~=~ \frac{1}{2} m \omega^2 x^2 ~.
\eeq
This turns out to be a rather special case because the ground state energy of 
the collective field theory can be found exactly. Since this is 
a static solution with $\theta =0$, we can use the principal part identity
\beq
\frac{P}{x-y} ~\frac{P}{x-z} ~+~ \frac{P}{y-z} ~\frac{P}{y-x} ~+~ 
\frac{P}{z-x} ~\frac{P}{z-y} ~=~ \pi^2 \delta (x-y) \delta (x-z) ~,
\eeq
to write (\ref{ene}) as a perfect square \cite{AND3}
\beq
E = \frac{\hbar^2}{2m} ~\int dx ~\rho ~\Bigl( ~\lambda \rho_H ~+~
\frac{\lambda -1}{2} \frac{\partial \rho}{\rho} ~-~ \frac{m \omega}{\hbar} 
x ~\Bigl)^2 ~ +~ \frac{\hbar \omega}{2} ~[ \lambda N^2 ~+~ (1-\lambda) N] ~.
\label{enesho}
\eeq
Thus if $\rho$ satisfies 
\beq
\lambda \rho_H ~+~ \frac{\lambda -1}{2} ~\frac{\partial \rho}{\rho} ~=~ 
\frac{m \omega}{\hbar}~ x ~,
\label{rhosho}
\eeq
then it minimizes (\ref{ene}) and is therefore a solution of the equation 
of motion (\ref{eom1}). Further, the ground state energy follows from 
(\ref{enesho}),
\beq
E_0 ~=~ \frac{\hbar \omega}{2} ~[ \lambda N^2 ~+~ (1-\lambda) N] ~.
\label{ensho}
\eeq
This is in fact the exact answer for the Hamiltonian (\ref{hamcsm}).

Eq. (\ref{rhosho}) for the density can be solved analytically only if 
$\lambda =0$ or $1$. We get
\bea
\rho ~&=&~ N ~\Bigl( \frac{m \omega}{4 \pi \hbar} \Bigr)^{1/2} ~\exp ~(~-m 
\omega x^2 /\hbar ~) \quad {\rm if} \quad \lambda =0 \nonumber \\
&=&~ \frac{mw}{\pi \hbar} ~\Bigl( ~\frac{2N \hbar}{m \omega} ~-~ x^2 ~ 
\Bigr)^{1/2} \quad {\rm if} \quad \lambda =1 ~.
\eea
We can show analytically that the collective field theory density has a 
Gaussian tail of the form
\beq
\rho ~\sim~ x^{2 \lambda N/(1-\lambda)} ~\exp ~\Bigl[~-~ \frac{m \omega 
x^2}{\hbar (1- \lambda)} ~\Bigr]
\eeq
if $\lambda < 1$, and has a sharp cutoff $x = \pm x_0$ beyond which $\rho$
vanishes if $\lambda \ge 1$. For large values of $N$, we can also show that 
the second term on the left hand side of (\ref{rhosho}) is generally much 
smaller than the first term; the scaling argument is indicated below. If we
ignore the second term altogether, we get the leading behavior of $\rho$ 
to be a semicircle for all nonzero values of $\lambda$,
\bea
\rho ~&=&~ \frac{mw}{\pi \hbar \lambda} ~\Bigl(x_0^2 ~-~ x^2 ~\Bigr)^{1/2} ~
\quad {\rm for} \quad \vert x \vert \leq x_0, \nonumber \\
&=&~ 0~ \quad {\rm for} \quad \vert x \vert ~> x_0 ~.
\label{semi}
\eea
Here $x_0$ is defined by 
\beq
x_0 ~=~ \Bigl( \frac{2N \hbar \lambda}{m\omega} \Bigr)^{1/2} ~.
\label{turn}
\eeq
The relations~(\ref{semi}) and (\ref{turn}) are identical to the Thomas-Fermi 
result obtained in \cite{SEN}, and $x_0$ is just the classical turning point. 
Note that the form of $\rho$ in (\ref{semi}) is essentially a statement 
of exlusion statistics for the CSM; the occupation number in each state in 
phase space $dx dp = 2 \pi \hbar$ is given by $1/ \lambda$. 
One should, however, be wary of using the expression 
(\ref{semi}) for the density $\rho(x)$. For example, if we indiscriminantly 
substitute this $\rho(x)$ in the static energy density functional 
\beq
E = \int dx ~V (x) ~\rho ~+~ \frac{\hbar^2}{2m} ~\int dx ~\Bigl[ ~
\frac{\pi^2 \lambda^2}{3} \rho^3 +~ \lambda (\lambda -1) \rho_H \partial 
\rho + \frac{(\lambda -1)^2}{4} ~\frac{(\partial \rho)^2}{\rho} ~\Bigr] \;,
\label{stat}
\eeq
the integrals with $V(x)$ and $\rho^3$ on the right-hand side together yield 
the correct $N^2$-dependent term in $E_0$, and the third integral gives the 
right $N$-dependent term (see Eq.~(\ref{ensho}), but the last integral 
involving $(\partial \rho)^2 / \rho$ diverges. It may be easily 
checked, however, that this divergent term goes like $N^0$, i.e., of order 
$1$ in the large-N expansion. Such terms in the expansion will be dropped.   

We should point out that the $1/N$ expansion of $\rho$ within the collective 
field theory cannot be taken too seriously; we recall the cautionary remarks 
following Eq. (\ref{rho}). For instance, the absence of a Gaussian tail 
if $\lambda \ge 1$ is an artifact of collective field theory. If $N$ is 
finite, the ground state of (\ref{hamcsm}) has a Gaussian tail for all values 
of $\lambda$; this can be seen from the exact expression
\beq
\Psi_0 ~[ x_i ] ~\sim ~ \prod_{i<j} ~\vert x_i - x_j \vert^{\lambda} ~\exp ~
[~- \frac{m \omega}{2\hbar} ~\sum_i ~x_i^2 ~] ~. 
\eeq
On the other hand, we can generally trust the next to leading term in the 
{\it energy} given by collective field theory. We have already seen this 
for the harmonic oscillator potential, where the $N^2$ and $N$-dependent 
terms both came out correctly. We will now show this for a somewhat more
general class of potentials.   

We formally define the first two terms in a $1/N$ expansion as follows. We 
assume that the ground state energy and chemical potential have expansions 
of the form
\bea
\mu ~&=&~ \mu^{(0)} ~+~ \mu^{(1)} ~, \nonumber \\
E_0 ~&=&~ E_0^{(0)} ~+~ E_0^{(1)} ~. 
\label{exp1}
\eea
where $E_0^{(1)}/E_0^{(0)}$ and $\mu^{(1)}/\mu^{(0)}$ are of order $1/N$.
For the density, we have an expansion of the form
\beq
\rho (x) ~=~ \rho^{(0)} (x) ~+~ \rho^{(1)} (x) ~.
\label{exp2}
\eeq
We will state the procedure for obtaining the expansion (\ref{exp2}) shortly.
But we can point out immediately that it is not a $1/N$ expansion for {\it 
all} positions $x$; although $\rho^{(1)} (x)/\rho^{(0)} (x)$ will {\it 
generally} be 
order $1/N$, that will not be true near the turning point $x_0$. The first 
term $\rho^{(0)}$ in (\ref{exp2}) is defined by considering only the 
$\rho^3$ and $V \rho$ terms in the energy functional (\ref{stat}) and hence
in Eq. (\ref{eom1}). The leading term in the chemical potential $\mu^{(0)}$
and the turning point $x_0$ are then fixed by the particle number constraint 
(\ref{cons}). Next, we define the term $\rho^{(1)}$ in (\ref{exp2}) 
by considering the $\rho^3$, $V \rho$ and $\rho_H \partial \rho$ terms in
(\ref{stat}) and the corresponding terms in (\ref{eom1}). Once again,
$\mu^{(1)}$ is fixed by the constraint (\ref{cons}). We will not go 
beyond the two leading terms in Eqs. (\ref{exp1}) and (\ref{exp2}), and will
therefore not need to consider the $(\partial \rho)^2 /\rho$ term in 
(\ref{stat}); this term actually diverges even more severely for $\rho^{(1)}$
than for $\rho^{(0)}$.

As a specific example of the $1/N$ expansion, let us now consider a confining 
potential of the power-law form
\beq
V (x) ~=~ \frac{\hbar^2}{2ma^2} ~\Bigl( ~\frac{\vert x \vert}{a} ~\Bigr)^p ~,
\eeq
where $p>0$, and $a$ is a measure of the width of the potential. (A
harmonic potential corresponds to the case $p=2$). For such a potential,
we can prove that the $V\rho + \rho^3$, $\rho_H \partial \rho$, and $(\partial
\rho)^2 /\rho$ terms in the energy functional (\ref{stat}) are successively
of higher order in $1/N$. To show this, let us define the dimensionless 
variables 
\bea
\tilde{x} ~&=&~ \frac{1}{N^{2/(p+2)}} ~\frac{x}{a} ~, \nonumber \\
\tilde{\rho} (\tilde{x}) ~&=&~ \frac{1}{N^{p/(p+2)}} ~a ~\rho (x) ~,
\eea
so that $\int d \tilde{x} \tilde{\rho} =1$. Then Eq. (\ref{stat}) takes the
form
\bea
E ~=~ \frac{\hbar^2}{2ma^2} ~N^{(3p+2)/(p+2)} ~\int d \tilde{x} ~\Bigl[ ~
\vert \tilde{x} \vert^p ~\tilde{\rho} ~&+&~ \frac{\pi^2 \lambda^2}{3} 
\tilde{\rho}^3 ~+~ \frac{\lambda (\lambda -1)}{N} \tilde{\rho}_H 
\tilde{\partial} \tilde{\rho} \nonumber \\
&+&~ \frac{(\lambda -1)^2}{4N^2} \frac{(\tilde{\partial} \tilde{\rho} 
)^2}{\tilde{\rho}} ~\Bigr] ~.
\eea
This justifies the form of the $1/N$ expansion given in the previous 
paragraph.  Following that procedure, the leading order terms in $\mu$ and 
$\rho$ are found to be
\bea
\mu^{(0)} ~&=&~ \frac{\hbar^2}{2ma^2} ~\Bigl ( \frac{x_0}{a} \Bigr)^p ~, 
\nonumber \\
\rho^{(0)} ~&=&~ \frac{2}{\pi \hbar \lambda} ~[~2m (\mu^{(0)} - V (x)~)~
]^{1/2} \quad {\rm if} \quad \vert x \vert \le x_0 ~.
\eea
{}From the constraint (\ref{cons}) we find
\bea
\Bigl( \frac{x_0}{a} \Bigr)^{1+p/2} ~&=&~ \frac{\pi \lambda N}{2 I_1} ~, 
\nonumber \\
I_1 ~&=&~ \int_0^1 ~dy ~\sqrt{1- y^p} ~=~ \frac{\Gamma (1 + \frac{1}{p} )~
\Gamma (\frac{3}{2})}{\Gamma (\frac{3}{2} + \frac{1}{p})} ~.
\eea
The leading terms $\rho^3$ and $V\rho$ in the energy (\ref{stat}) then give
\bea
E_0^{(0)} ~&=&~ \frac{2}{3\pi \hbar \lambda} ~\int_0^{x_0} ~dx ~[~ 2m 
(\mu^{(0)} - V(x) ~]^{1/2} ~(\mu^{(0)} + 2 V(x) ) \nonumber \\
&=&~ \frac{\hbar^2}{2ma^2} ~\frac{p+2}{3p+2} ~\Bigl( \frac{\pi \lambda}{2 
I_1} \Bigr)^{2p/(p+2)} ~N^{(3p+2)/(p+2)} ~.
\label{e00}
\eea
We now go to next order in $1/N$ by including the terms in (\ref{eom1})
and (\ref{stat}) which contain the Hilbert transform $\rho_H$; we get
\beq
\frac{\pi^2 \hbar^2 \lambda^2}{m} ~\rho^{(0)} \rho^{(1)} ~-~ \frac{\hbar^2 
\lambda (\lambda -1)}{m} ~\partial \rho_H^{(0)} ~=~ \mu^{(1)} ~.
\label{exp3}
\eeq
The structure of (\ref{exp3}) shows that $\rho^{(1)}$ must be taken to be zero
outside the turning point $x_0$, just like $\rho^{(0)}$. Since $\rho^{(0)}$ is
normalized to $N$, we fix $\mu^{(1)}$ using the constraint
\beq
\int_{-x_0}^{x_0} ~dx ~\rho^{(1)} ~=~ 0 ~,
\eeq
and we then determine $E_0^{(1)}$ by including the 
$\rho_H \partial \rho$ term in 
(\ref{stat}). Interestingly, we find that $\rho^{(1)}  =0$ for all $\lambda$
for the simple harmonic case ($p=2$); that is why we get the energy correct 
to order $N$ by just substituting $\rho^{(0)}$ in (\ref{stat}). However 
$\rho^{(1)}$ is not zero for a general value of $p$. We will omit the final 
expression for $E_0^{(1)}$ for general $\lambda$,
and will now specialize to $\lambda =1$ where we can compare with the results
of a WKB approximation. (Although this is a free fermion theory, it is an 
interacting bosonic theory. Hence agreement at $\lambda =1$ is a nontrivial 
check of collective field theory). We find that both $\mu^{(1)}$ and 
$\rho^{(1)} (x)$ are zero for $\lambda =1$. Hence there is {\it no} correction 
to $E_0$ at the next order after (\ref{e00}), i.e., at order $N^{2p/(p+2)}$. 
We will now show that this result agrees with WKB.

If $e_n$ denote the single-particle energy levels
obtained by solving the Schr\"{o}dinger equation in the potential
$V (x)$, then the exact ground state energy at $\lambda =1$ is given by
\beq
E_0 ~=~ \sum_{n=0}^{N-1} ~e_n ~.
\label{e0wkb}
\eeq
For large $n$, we can obtain the two leading order terms in $e_n$ using the 
WKB formula
\beq
\int_{-x_n}^{x_n} ~dx ~[~ 2m ~(~e_n - V(x)~)~]^{1/2} ~=~ \Bigl(~  n+
\frac{1}{2} ~\Bigr) ~\pi \hbar ~,
\eeq
where $x_n$ denotes the classical turning point for energy $e_n$. We thus 
obtain the expansion
\beq
e_n ~=~ \frac{\hbar^2}{2ma^2} ~\Bigl( \frac{\pi n}{2 I_1} \Bigr)^{2p/(p+2)} ~
[~ 1 ~-~ \frac{p}{(p+2)n} ~+~ \cdot \cdot \cdot ~] ~.
\eeq
On substituting this in (\ref{e0wkb}), we find that $E_0$ is indeed given by 
(\ref{e00}) and that there is no correction to the next order in $1/N$.

\vskip .5 true cm
\centerline{4. SMALL AMPLITUDE WAVES, CORRELATION FUNCTIONS, AND}
\centerline{SPECIFIC HEAT}
\vskip .5 true cm

In this Section, we will study the small amplitude density fluctuations
about an uniform background density $\rho_0$. We will show that these
exhaust the low-energy excitations upto some order, both in a sum rule and 
in the low-temperature specific heat.

For an uniform density $\rho_0$, the chemical potential is given by 
(\ref{mu}) or (\ref{eom1}) to be
\beq
\mu ~=~ \frac{\pi^2 \hbar^2 \lambda^2 \rho_0^2}{2m} ~.
\eeq
Let us now study (\ref{eom1}) and (\ref{eom2}) to linear order in an 
amplitude $a << 1$. We assume
\bea
\rho ~&=&~ \rho_0 ~+~ a \rho_0 ~\cos (kx- \omega t) ~, \nonumber \\
\theta ~&=&~ a ~\frac{m \omega}{\hbar k^2} ~\sin (kx - \omega t)~,
\label{sou}
\eea
where $k$ denotes the wave number; the second equation in 
(\ref{sou}) follows from the first due to the equation of continuity
(\ref{eom2}). Eq. (\ref{eom1}) then yields the dispersion relation
\beq
\omega_k ~=~ \vert ~\frac{\pi \hbar \lambda \rho_0 \vert k \vert }{m} ~-~ 
\frac{(\lambda -1) \hbar k^2}{2m} ~\vert ~.
\label{disp}
\eeq
The sound velocity is given by the group velocity $\partial \omega / \partial
k$ at $k=0$; the result agrees with the exact value given in (\ref{vs}).
Note that (\ref{disp}) gives the correct single-particle dispersion 
for $\lambda =0$, as expected for a free boson theory.

We see that the dispersion (\ref{disp}) vanishes not only at $k=0$, but 
also at
\beq
\vert k \vert ~=~ \frac{2 \pi \lambda \rho_0}{\lambda -1} 
\label{k0}
\eeq
if $\lambda > 1$. 
However the latter point where $\omega$ vanishes seems to be an artifact of 
collective field theory; it does not agree with known results. For instance, 
reference \cite{MUC} defines a dispersion relation called the "Feynman 
spectrum" as follows. Consider the dynamical correlation function and its 
Fourier transform
\bea
G(x,t) ~&=&~ \langle ~\rho (x,t) \rho (0,0) ~\rangle ~-~ \rho_0^2 ~, 
\nonumber \\
S(k,\omega) ~&=&~ \frac{1}{2\pi \rho_0} ~\int dx \int dt ~G(x,t) ~
e^{-i (kx-\omega t)} ~.
\label{corr}
\eea
$S(k,\omega)$ can be represented in terms of all the states of the system
$\vert n \rangle$ with energies $E_n$ as
\bea
S(k,\omega) ~&=&~ \frac{1}{N} ~\sum_{n} ~\vert \langle n \vert \rho_k \vert
0 \rangle \vert^2 ~\delta (\hbar \omega - E_n + E_0) ~, \nonumber \\
\rho_k ~&=&~ \sum_{n=1}^N ~e^{-ikx_n} ~.
\eea
One can then define various moments of $S(k,\omega)$ as
\beq
I_n (k) ~=~ \int d\omega ~\omega^n ~S(k,\omega) ~,
\eeq
where $I_1 (k) = k^2 /2m$. The Feynman dispersion is defined as 
\beq
\omega_F (k) ~=~ \frac{I_1 (k)}{I_0 (k)} ~.
\eeq
Now this dispersion is known from random matrix theory \cite{SUT1,MEH} for 
three special values of $\lambda = 1/2, 1$ and $2$. For these three
values, it is found \cite{MUC} that $\omega_F (k)$ agrees with 
(\ref{disp}) upto second order for $k$ close to $0$. We therefore believe
that $\omega_F (k)$ is given by (\ref{disp}) to order $k^2$ for {\it all}
values of $\lambda$. However, the agreement between $\omega_F (k)$ and
(\ref{disp}) does not persist to higher orders in $k$ even for the three
special values of $\lambda$; in particular, $\omega_F (k)$ does not
vanish at any nonzero values of $k$, although it does have a roton-like 
minimum at $\vert k \vert = 2 \pi \rho_0$ for $\lambda > 1$ \cite{MUC}.
This discrepancy between $\omega_F (k)$ and (\ref{disp}) seems to indicate 
that collective field theory cannot be trusted for large values of the 
wavenumber; it seems to work only upto order $k^2$.

Our statement that the low-energy dispersion is correctly given by
(\ref{disp}) upto $k^2$ near $k=0$ also agrees with the known low-temperature 
specific heat of the CSM to second order in the temperature $T$ \cite{ISA2}. 
We can compute the free energy per unit length $f$ from (\ref{disp}) taking
the sound modes to have zero chemical potential. Thus
\beq
\beta f ~=~ \int_{-\infty}^{\infty} ~\frac{dk}{2\pi} ~\ln ~(~1~-~ e^{-
\beta \hbar \omega_k} ~) ~,
\eeq
where $\beta = 1/k_B T$. After evaluating this, we can obtain the specific 
heat per unit length $C_V = -T \partial^2 f / \partial T^2$ to second order
in $T$. We find 
\bea
C_V ~&=&~ \frac{\pi k_B^2 T}{3 \hbar v_s} ~+~ \frac{6 \zeta (3) (\lambda -
1)}{\pi}~ \frac{k_B^3 T^2}{m \hbar v_s^3} ~, \nonumber \\
\zeta (3) ~&=&~ \sum_{n=1}^{\infty} ~\frac{1}{n^3} ~.
\label{cv}
\eea
This agrees with the result in reference \cite{ISA2}.

We note that the linear terms in (\ref{disp}) and (\ref{cv}) are typical of 
a system whose low-energy and long-wavelength excitations are governed by a 
conformal field theory \cite{FUK}. The quadratic terms in those two equations 
indicate deviations from conformal field theory which start appearing when 
the wavelength is no longer much longer than the typical particle spacing 
$1/ \rho_0$.

Finally, it may be useful to see what we get if we compute the correlation 
function $G(x,t)$ defined in (\ref{corr}) by {\it quantizing} the collective 
field theory. Using the commutation relation (\ref{comm}) and the equations of 
motion (\ref{eom1}) and (\ref{eom2}) to linear order, we find that $\rho$ and 
$\theta$ have the following second quantized expressions,
\bea
\rho ~&=&~ \rho_0 ~+~ \int_{-\infty}^{\infty} ~\frac{dk}{2 \pi} ~f_k ~\Bigl[ ~
a_k ~e^{i(kx - \omega_k t)} ~+~ a_k^{\dag} ~e^{-i(kx - \omega_k t)} ~\Bigr] ~,
\nonumber \\
\theta ~&=&~ \frac{i}{2} ~\int_{-\infty}^{\infty} ~\frac{dk}{2 \pi} ~
\frac{1}{f_k} ~ \Bigl[ ~-~ a_k ~e^{i(kx - \omega_k t)} ~+~ a_k^{\dag} ~
e^{-i(kx - \omega_k t)} ~ \Bigr] ~, \nonumber \\
f_k ~&=&~ \Bigl( ~\frac{\hbar \rho_0 k^2}{2 m \omega_k} ~\Bigr)^{1/2} ~,
\eea
where
\beq
[ ~a_k ~,~ a_{k^{\prime}}^{\dag} ~] ~=~ 2 \pi ~\delta (k-k^{\prime} ~) ~.
\eeq
{}From this we find that 
\beq
G(x,t) ~=~ \frac{\hbar \rho_0}{2m} ~\int_{-\infty}^{\infty} ~\frac{dk}{2 \pi}~
\frac{k^2}{\omega_k} ~e^{i(kx -\omega_k t)} ~.
\label{gxt1}
\eeq
If we now use the collective field theory dispersion (\ref{disp}), the 
integral will diverge at the nonzero values of $\vert k \vert$ where $\omega$ 
vanishes.  We will therefore assume, as stated above, that (\ref{disp}) 
can only be trusted in the region near $k=0$. The asymptotic form of 
$G(x,t)$ at large values of $x \pm 
v_s t$ only gets a contribution from that region in the integral (\ref{gxt1});
further, only the linear term in the dispersion (\ref{disp}) is required to
derive the asymptotic expression. We then find that
\beq
G(x,t) ~\sim ~ - ~\frac{1}{4 \pi^2 \lambda} ~\Bigl[ ~\frac{1}{(x - v_s t)^2} ~
+~ \frac{1}{(x + v_s t)^2} ~\Bigr] ~.
\label{gxt2}
\eeq
This agrees with the leading nonoscillating term in the exact expression given 
in reference \cite{HA}. However the exact expression also has oscillating
terms; in fact, such a term dominates over (\ref{gxt2}) if $\lambda > 1$. The 
fact that collective field theory is unable to reproduce these oscillating 
terms clearly shows its limitation.

\vskip .5 true cm
\centerline{5. LARGE AMPLITUDE WAVES}
\vskip .5 true cm

Following a method given in reference \cite{POL2}, we can find exact solutions
which describe waves with arbitrary amplitude. We will consider the cases
$\lambda > 1$ and $0 < \lambda < 1$ separately.

For $\lambda > 1$, the solutions are given by
\bea
\rho (x,t) ~&=&~ \frac{(\lambda -1)k}{2 \pi \lambda} ~\Bigl[ ~c~ +~
\frac{\sinh \alpha}{\cosh \alpha - \cos k (x -v t)} ~\Bigr] ~, \nonumber \\
\theta (x,t) ~&=&~ \frac{mv}{\hbar} ~\Bigl[~- ~(c+1) ~\int_{-\infty}^x ~
dy ~\frac{\cosh \alpha - \cos k (y-vt)}{c \cosh \alpha + \sinh \alpha - c 
\cos k (y-vt)} \nonumber \\
&& \quad \quad \quad +~ x ~-~ \frac{1}{2} ~v t ~\Bigr] ~, 
\label{wave}
\eea
We choose the wavenumber $k$ and the parameter $\alpha$ to be positive.
The phase velocity $v$ in (\ref{wave}) satisfies
\beq
v^2 ~=~ \frac{(\lambda -1)^2 \hbar^2 k^2 c^2}{4 m^2 (c+1)^2} ~(c^2 + 2
c \coth \alpha + 1) ~, 
\label{vel}
\eeq
where
\beq
c ~\ge ~\frac{1- \cosh \alpha}{\sinh \alpha} ~.
\eeq
The average density for this solution is found to be
\beq
\rho_0 ~=~ \frac{(\lambda -1)k}{2 \pi \lambda} ~(c+1) ~.
\label{averho}
\eeq
We can define the dimensionless amplitude $a$ of the wave to be the 
fractional difference between the maximum and minimum densities. Thus
\beq
a ~=~ \frac{\rho_{max} - \rho_{min}}{\rho_{max} + \rho_{min}} ~=~ 
\frac{1}{c \sinh \alpha + \cosh \alpha} ~.
\eeq

For $\lambda < 1$, the solutions are again given by Eqs. (\ref{wave}) and 
(\ref{vel}), but
\beq
c ~\le ~\frac{-1- \cosh \alpha}{\sinh \alpha} ~.
\eeq
The average density for this solution is the same as in Eq. (\ref{averho}),
while the amplitude $a$ is
\beq
a ~=~ -~ \frac{1}{c \sinh \alpha + \cosh \alpha} ~.
\eeq

The solutions above are characterized by three independent parameters
which may be considered to be the average density $\rho_0$, the wavenumber 
$k$, and the amplitude $a$. If we hold $\rho_0$ and $k$ fixed and let 
$\alpha \rightarrow \infty$, we recover the small amplitude waves discussed in 
the previous Section. Let us now look at the conditions under which the 
frequency $\omega = \vert v k \vert$ can vanish; this corresponds to 
stationary waves. We can see from Eq. (\ref{vel}) that $\omega$
vanishes if $k=0$ or if $c=0$; the latter is allowed if $\lambda > 1$
in which case $k$ satisfies (\ref{k0}). These two conditions for $\omega =0$
are therefore the same as those found for the sound modes in Section 4. 
However we now see that $\omega$ also vanishes if 
\beq
c ~=~ \frac{1 - \cosh \alpha}{\sinh \alpha}
\label{c1}
\eeq
for $\lambda >1$, or if
\beq
c ~=~ \frac{-1 - \cosh \alpha}{\sinh \alpha}
\label{c2}
\eeq
for $\lambda <1$. In these two cases, we get staionary waves with the 
largest possible amplitude $a=1$ since $\rho_{min} =0$.

The interpretation of all these large amplitude waves, including the new kinds 
of stationary waves in (\ref{c1}) and (\ref{c2}), in terms of the exact 
solutions of the CSM model remains an open question. It is possible that some 
of the solutions obtained here are peculiar to the collective field theory
and do not correspond to anything in the CSM. For instance, there are no 
exact solutions of the CSM which have arbitrary nonzero values of $k$ with 
$\omega =0$.

In concluding, we would like to mention that the density waves studied 
in this Section and in the previous Section were known earlier \cite{POL2} 
for {\it large} $\lambda$. In addition, the stationary waves (\ref{c2}) for 
$\lambda < 1$ were found in reference \cite{AND1}. Our own results describe 
both stationary and moving waves, and are valid for all values of $\lambda$.

\vskip .5 true cm
\centerline{6. SINGLE SOLITONS}
\vskip .5 true cm

We will now describe the single soliton solutions of the collective field
theory \cite{POL2,AND1}. Starting from the large amplitude waves in Section 
5, we can find these solutions for any $\lambda$ different from $0$ and $1$ 
as follows. In Eq. (\ref{wave}), we take the limit $k , \alpha \rightarrow 0$ 
keeping $\alpha /k =b$ fixed. Simultaneously, we let $c \rightarrow \infty$ 
for $\lambda > 1$ or $- \infty$ for $\lambda < 1$, keeping $\rho_0$ fixed 
according to (\ref{averho}). Since the wavelength $2\pi /k \rightarrow 
\infty$, we obtain a solution describing an isolated lump. We find the 
following expressions for $\rho$ and $\theta$ in terms of the width $b$ and 
velocity $v$ of the soliton.
\bea
\rho (x,t) ~&=&~ \rho_0 ~+~ \frac{\lambda -1}{\pi \lambda} ~\frac{b}{(x-vt)^2 
+ b^2} ~, \nonumber \\
\theta (x,t) ~&=&~ \pm ~(\lambda -1) ~\tan^{-1} ~\Bigl( ~\frac{x-vt}{b~
\eta} ~\Bigr) ~-~ \frac{1}{2} ~ \frac{m v^2 t}{\hbar} ~, \nonumber \\
{\rm where} \quad v ~&=&~ \pm ~\frac{\pi \hbar \lambda \rho_0}{m} ~\eta ~,
\nonumber \\
\eta ~&=&~ \Bigl[ ~ 1 ~+~ \frac{\lambda -1}{\pi \lambda \rho_0 b} ~
\Bigr]^{1/2} ~.
\label{sol}
\eea
Thus the velocity and width are related to each other. From Eq. 
(\ref{vs}) we see that $\vert v \vert \ge v_s$ and any value of $b 
\rho_0$ is allowed for $\lambda > 1$. For $\lambda < 1$, we must have 
$b \rho_0 \ge (1- \lambda)/\pi \lambda$; then $\vert v \vert \le v_s$. 
These ranges of velocity agree with the exact results known for the particle 
and hole respectively, as discussed after Eq. (\ref{enh}). The identification
with particle for $\lambda > 1$ and hole for $\lambda < 1$ may be justified
by considering the {\it sign} of the integrated density for the soliton, 
\beq
\int dx ~[~ \rho (x,t) ~-~ \rho_0 ~] ~=~ \frac{\lambda - 1}{\lambda} ~.
\eeq
The magnitude of this number is generally not an integer; the physical meaning 
of this is not clear to us. It is interesting to note at this point that if 
we perform the scaling (\ref{red}) which eliminates $\lambda$, then 
the redefined soliton number is $1$ for all $\lambda$. Hence (\ref{sol})
describes a one particle solution in the redefined theory.

The momentum and energy (obtained after subtracting the background value)
of the soliton (\ref{sol}) are given by Eqs. (\ref{mom}) and (\ref{ene}).
\bea
P ~&=&~ \frac{\lambda -1}{\lambda} ~m v ~, \nonumber \\
E ~&=&~ \frac{\lambda -1}{\lambda} ~( ~\frac{1}{2} ~m v^2 ~-~ \mu ~) ~.
\eea
This dispersion relation does {\it not} agree with the exact dispersion 
relations given in (\ref{enp}) and (\ref{enh}). It therefore seems that the 
interpretation of solitons as particles or holes has some difficulties which
need to be resolved. 

It is interesting to observe that we can also go in the opposite direction
and recover the large amplitude waves by superposing a number of single
soliton solutions in a periodic way \cite{POL2}. For this purpose, it is 
useful to recall that
\beq
2 \alpha ~\sum_{n=-\infty}^{\infty} ~\frac{1}{(2\pi n + kx)^2 + \alpha^2} ~=~
\frac{\sinh \alpha}{\cosh \alpha - \cos ~(kx)} ~.
\eeq

It is worth noting that for $\lambda > 1$, there is no lower bound on
$\rho_0$; in particular, we can set $\rho_0 =0$. We then get a new solution 
corresponding to a stationary and isolated soliton with no background density.
This may also be seen for the static case from Eq. (\ref{rhosho}) with no 
external harmonic potential,
\beq
\lambda \rho_H ~+~ \frac{\lambda -1}{2} ~\frac{\partial \rho}{\rho} ~=~ 0\;.
\label{solo}
\eeq
The solution of this is found to be 
\beq
\rho (x) ~=~ \frac{\lambda -1}{\pi \lambda} ~\frac{b}{x^2 
+b^2} ~, 
\label{kite}
\eeq
with eigenvalue $E_0=0$. 
We may now boost this solution using (\ref{gal}). The general solution 
is therefore
\bea
\rho (x,t) ~&=&~ \frac{\lambda -1}{\pi \lambda} ~\frac{b}{(x-vt)^2 
+b^2} ~, \nonumber \\
\theta (x,t) ~&=&~ \frac{m v}{\hbar} ~\Bigl[ ~ x ~-~ \frac{1}{2} v t ~
\Bigr] ~,
\label{sol2}
\eea
where the width and velocity are now independent parameters. The particle 
number, momentum and energy of this soliton are given by
\bea
N ~&=&~ \frac{\lambda - 1}{\lambda} ~, \nonumber \\
P ~&=&~ \frac{\lambda - 1}{\lambda} ~ m v ~, \nonumber \\
E ~&=&~ \frac{\lambda - 1}{\lambda} ~\frac{mv^2}{2} ~. 
\eea
Note that if we had used the scaled variables given by Eq. (\ref{red}), 
the normalized $dx \int \tilde{\rho} (x) =1$, with the momentum and energy 
like a classical particle. Even though it is an exact solution of the 
collective field theory, it not a genuine solution of the CSM. This is because 
the collective field theory is meaningful only for large $N$. 

Finally, we note that we may rewrite Eq. (\ref{solo}) for the scaled density 
$\tilde{\rho}$ as 
\beq
\frac{1}{2} ~\frac{\partial^2 \tilde{\rho}}{\partial x^2} ~+~ 
\frac{\partial}{\partial x} \left( ~\tilde{\rho}\tilde{\rho}_H ~\right)~=~0 ~.
\eeq
Formally, this equation has the same form as the steady-state Coulomb gas 
model of Dyson \cite{DYS}. In the diffusion problem, it is known as the 
Smoluchowski equation with a singular kernel, and describes the Brownian 
motion of a particle immersed in a fluid, with friction-limited velocity. 
A description of this equation is given by Andersen and 
Oppenheim \cite{ANDE}. The analogous single-soliton solution (\ref{kite}) 
of the equation for the diffusion problem was obtained by Satsuma and 
Mimura \cite{SAT}. These authors also found the soliton with the hyperbolic 
kernel, and the periodic solution appropriate for the Sutherland Hamiltonian 
on a circle \cite{SUT1}. 

\vskip .5 true cm
\centerline{7. DISCUSSION}
\vskip .5 true cm

We have seen that collective field theory is a powerful technique from which 
many properties of the CSM can be derived without having to solve the 
$N$-particle Schr\"{o}dinger equation (\ref{hamcsm}). We can consider other
applications of collective field theory. For instance, it should be possible 
to solve for the low-energy excitations of the CSM in a slowly time-varying 
harmonic potential. Analogous calculations have been performed for a trapped 
Bose-Einstein condensate in the three-dimensional problem \cite{CAS}.

It is evident that there are several issues which are either not clear or 
beyond the reach of collective field theory. We list some of these problems 
below; they are of course related to each other. 

\noindent
(a) The dispersion relation of small amplitude waves whose wavelengths are 
comparable to the average particle spacing remains unknown. The difficulties
mentioned in Section 4 seem to suggest that the collective field theory 
discussed in this paper is not complete; perhaps one needs higher derivative 
terms in the energy functional (\ref{ene}) to obtain better results at
short wavelengths. 

\noindent
(b) The interpretation of the large amplitude waves and soliton solutions 
is not clear. The ideal way to resolve these difficulties would be to set 
up a precise correspondence between the collective excitations and the known 
solutions of the Schr\"{o}dinger equation. Refs. \cite{SUT2,POL2} make some 
suggestions in this direction, but a quantitative mapping is still missing. 

\noindent 
(c) It is not clear why we only get exact soliton solutions which correspond 
to particles for $\lambda > 1$ and holes for $\lambda <1$. It would be 
desirable to complete the story by finding, perhaps numerically, solutions 
corresponding to particles for $\lambda < 1$ and holes for $\lambda > 1$.

\noindent
(d) Finally, it would be very interesting to quantize the collective field
theory and study it more carefully than we have done in Section 4. This 
may lead to an alternative way of deriving the oscillating terms in the 
dynamical correlation functions. 

\vskip .5 true cm

RKB would like to thank Akira Suzuki for studying the single soliton solutions
in the early stages of this work, and M. V. N. Murthy for discussions on the
collective field theory. DS thanks the Department of Physics and Astronomy, 
McMaster University for its hospitality during the course of this work. This 
research was supported by the Natural Sciences and Engineering Research 
Council of Canada.

%\newpage


\begin{thebibliography}{99}
\vskip .5 true cm

\bibitem{CAL} F. Calogero, {\it J. Math. Phys.} {\bf 10} (1969), 2191, 2197.

\bibitem{SUT1} B. Sutherland, {\it J. Math. Phys.} {\bf 12} (1971), 246, 251; 
{\it Phys. Rev. A} {\bf 4} (1971), 2019; {\it Phys. Rev. A} {\bf 5} (1972), 
1372; {\it Phys. Rev. B} {\bf 45} (1992), 907.

\bibitem{MOS} J. Moser, {\it Adv. Math.} {\bf 16} (1975), 197; F. Calogero, 
{\it Lett. Nuovo Cimento} {\bf 13} (1975), 411.

\bibitem{MEH} M. L. Mehta, "Random Matrices," Academic Press, New York, 1991;
B. D. Simons, P. A. Lee and B. L. Altshuler, {\it Phys. Rev. Lett.} {\bf 72}
(1994), 64.

\bibitem{HAL1} F. D. M. Haldane, {\it Phys. Rev. Lett.} {\bf 60} (1988), 635; 
B. S. Shastry, {\it Phys. Rev. Lett.} {\bf 60} (1988), 639.

\bibitem{HAL2} F. D. M. Haldane, {\it Phys. Rev. Lett.} {\bf 67} (1991), 937. 

\bibitem{MUR} M. V. N. Murthy and R. Shankar, {\it Phys. Rev. Lett.} {\bf 72}
(1994), 3629; {\it Phys. Rev. Lett.} {\bf 73} (1994), 3331.

\bibitem{DAS} A. Dasnieres de Veigy and S. Ouvry, {\it Phys. Rev. Lett.} {\bf 
72} (1994), 600; {\it Mod. Phys. Lett. A} {\bf 10} (1995), 1.

\bibitem{WU} Y.-S. Wu, {\it Phys. Rev. Lett.} {\bf 73} (1994), 922; D. 
Bernard and Y.-S. Wu, Proceedings of the 6th Nankai workshop on "New 
developments of integrable systems and long-ranged interaction models", eds. 
M. L. Ge and Y.-S. Wu, World Scientific, Singapore, 1995.

\bibitem{ISA1} S. B. Isakov, {\it Phys. Rev. Lett.} {\bf 73} (1994), 2150; 
{\it Int. J. Mod. Phys. A} {\bf 9} (1994), 2563; {\it Mod. Phys. Lett. B} 
{\bf 8} (1994), 319.

\bibitem{ISA2} S. B. Isakov, D. P. Arovas, J. Myrheim and A. P. Polychronakos, 
{\it Phys. Lett. A} {\bf  212} (1996), 299; C. Nayak and F. Wilczek, {\it 
Phys. Rev. Lett.} {\bf 73} (1994), 2740. 

\bibitem{SEN} D. Sen and R. K. Bhaduri, {\it Phys. Rev. Lett.} {\bf 74} 
(1995), 3912.

\bibitem{FUK} N. Kawakami and S.-K. Yang, {\it Phys. Rev. Lett.} {\bf 67}, 
(1991) 2493; A. G. Izergin, V. E. Korepin and N. Yu. Reshetikhin, {\it J. 
Phys. A} {\bf 22} (1989), 2615; T. Fukui and N. Kawakami, {\it Phys. Rev. B}
{\bf 51} (1995), 5239; {\it J. Phys. A} {\bf 28} (1995), 6027. 

\bibitem{YU} N. Kawakami, {\it Phys. Rev. Lett.} {\bf 71} (1993), 275; Y. Yu 
and Z. Zhu, Academia Sinica (Beijing) preprint (1996).

\bibitem{CHE} H. H. Chen, Y. C. Lee and N. R. Pereira, {\it Phys. Fluids}
{\bf 22} (1979), 187.

\bibitem{POL1} A. P. Polychronakos, {\it Phys. Rev. Lett.} {\bf 69} (1992), 
703; B. Sutherland and B. S. Shastry, {\it Phys. Rev. Lett.} {\bf 71} (1993), 
5. 

\bibitem{SUT2} B. Sutherland and J. Campbell, {\it Phys. Rev. B} {\bf 50} 
(1994), 888.

\bibitem{POL2} A. P. Polychronakos, {\it Phys. Rev. Lett.} {\bf 74} (1995), 
5153.

\bibitem{AND1} I. Andri\'{c}, V. Bardek and L. Jonke, {\it Phys. Lett. B} 
{\bf 357} (1995), 374; {\it J. Phys. A} {\bf 30} (1997), 717.

\bibitem{BEC} E. M. Lifshitz and L. P. Pitaevskii, "Statistical Physics", 
Part 2, chapter 3, Pergamon Press, Oxford, 1980; L. P. Pitaevskii, {\it Zh. 
Eksp. Teor. Fiz.} {\bf 40} (1961), 646; E. P. Gross, {\it Nuovo Cimento} {\bf 
20} (1961), 454.  

\bibitem{JEV1} A. Jevicki and B. Sakita, {\it Nucl. Phys. B} {\bf 165} (1980), 
511.

\bibitem{AND2} I. Andri\'{c}, A. Jevicki and H. Levine, {\it Nucl. Phys. B}
{\bf 215} (1983), 307.

\bibitem{AND3} I. Andri\'{c} and V. Bardek, {\it J. Phys. A} {\bf 21} (1988), 
2847.

\bibitem{JEV2} A. Jevicki, {\it Nucl. Phys. B} {\bf 376} (1992), 75.

\bibitem{MUC} E. R. Mucciolo, B. S. Shastry, B. D. Simons and B. L. 
Altshuler, {\it Phys. Rev. B} {\bf 49} (1994), 15197.

\bibitem{SKY} T. H. R. Skyrme, {\it Phil. Mag.} {\bf 1} (1956), 1043; P. 
Hohenberg and W. Kohn, {\it Phys. Rev.} {\bf 136} (1964), B 864; W. Kohn and 
L. J. Sham, {\it Phys. Rev. A} {\bf 137} (1965), 1697; {\bf 140} (1965), 1133.

\bibitem{HIL} "Table of Integral Transforms", Vol. II, Bateman Manuscript
Project, ed. A. Erd\'{e}lyi, chapter XV, Mc-Graw Hill, New York, 1954.

\bibitem{HA} Z. N. C. Ha, {\it Nucl. Phys. B} {\bf 435} (1995), 604; J. A. 
Minahan and A. P. Polychronakos, {\it Phys. Rev. B} {\bf 50} (1994), 4236.

\bibitem{MAD} E. Madelung, {\it Z. Phys.} {\bf 40} (1928), 332.
 
\bibitem{BOH} D. Bohm, {\it Phys. Rev.} {\bf 85} (1952), 166; {\bf 85} (1952), 
180.
 
\bibitem{DYS} F. J. Dyson, {\it J. Math. Phys.} {\bf 3} (1962), 1191. 

\bibitem{ANDE} H. C. Andersen and I. Oppenheim, {\it J. Math. Phys.} {\bf 
4} (1963), 1367.

\bibitem{SAT} J. Satsuma and M. Mimura, {\it J. Phys. Soc. Jpn.} {\bf 54}
(1985), 894.

\bibitem{CAS} Y. Castin and R. Dum, {\it Phys. Rev. Lett.} {\bf 77} (1996), 
5315.

\end{thebibliography}
\end{document}